\documentclass[doublecol]{epl2}
\title{Reentrant phase transitions and defensive alliances in social dilemmas with informed strategies}
\shorttitle{Reentrant phase transitions and defensive alliances in social dilemmas with informed strategies}

\author{Attila Szolnoki,\inst{1} Matja{\v z} Perc,\inst{2,3,4}}
\shortauthor{Szolnoki and Perc}

\institute{\inst{1}Institute of Technical Physics and Materials Science, Centre for Energy Research, Hungarian Academy of Sciences, P.O. Box 49, H-1525 Budapest, Hungary\\
\inst{2}Faculty of Natural Sciences and Mathematics, University of Maribor, Koro{\v s}ka cesta 160, SI-2000 Maribor, Slovenia\\
\inst{3}Department of Physics, Faculty of Sciences, King Abdulaziz University, Jeddah, Saudi Arabia\\
\inst{4}CAMTP -- Center for Applied Mathematics and Theoretical Physics, University of Maribor, Krekova 2, SI-2000 Maribor, Slovenia}


\abstract{Knowing the strategy of an opponent in a competitive environment conveys obvious evolutionary advantages. But this information is costly, and the benefit of being informed may not necessarily offset the additional cost.
Here we introduce social dilemmas with informed strategies, and we show that this gives rise to two cyclically dominant triplets that form defensive alliances. The stability of these two alliances is determined by the rotation velocity of the strategies within each triplet. A weaker strategy in a faster rotating triplet can thus overcome an individually stronger competitor. Fascinating spatial patterns favor the dominance of a single defensive alliance, but enable also the stable coexistence of both defensive alliances in very narrow regions of the parameter space. A continuous reentrant phase transition reveals before unseen complexity behind the stability of strategic alliances in evolutionary social dilemmas.}

\begin{document}

\maketitle

Successful evolution of cooperation in the realm of social dilemmas is a grand challenge that continues to attract research across the social and natural sciences \cite{szabo_pr07, santos_jtb12, nowak_jtb12, perc_bs10, rand_tcs13, pacheco_plrev14}. When individuals are torn between what is best for them and what is best for their society, cooperation can quickly come to play second fiddle to the pursuit of short-term personal benefits. Although studies in evolutionary game theory \cite{mestertong_01, nowak_06, sigmund_10,lugo_srep15,wu_zx_pre14} have revealed fundamental rules that promote cooperation \cite{nowak_s06}, the chasm behind the Darwinian ``only the fittest survive'' and the abundance of cooperation in human and animal societies remains overwhelming.

The availability of information has been the holy grail in modern game theory since its inception \cite{neumann_44}. If one knew the rules of the game and what strategy the opponent will play, then the best response would be guaranteed \cite{macy_pnas02,rustagi_s10,szolnoki_pre12}. In repeated evolutionary settings with many players and under the reasonable assumption that information about the opponents is not free, however, the prospects are much less clear. In fact, recent human experiments have revealed that, contrary to expectations, costly information about the neighbors can impair the evolution of cooperation in social dilemmas \cite{antonioni_pone14}. Important unanswered questions are thus: (i) Under which conditions are informed strategies --- players that invest into learning the strategies of other players --- evolutionary stable? (ii) What are the properties of these stable solutions? and (iii) Which mechanisms are responsible for their stability?

In this letter, we answer these questions by introducing social dilemmas with informed strategies. The following matrix describes the payoffs of the four competing strategies:\\
\begin{center}
\begin{tabular}{r|c c c c}
 & $C$ & $D$ & $I_C$ & $I_D$\\
\hline
$C$ & $R$ & $S$ & $R$ & $S$\\
$D$ & $T$ & $P$ & $0$ & $P$\\
$I_C$ & \,\,\, $R-\epsilon$ \,\,\,& $0-\epsilon$ \,\,\,& $R-\epsilon$ \,\,\,& $0-\epsilon$\\
$I_D$ & \,\,\, $T-\epsilon$ \,\,\,& $0-\epsilon$ \,\,\,& $0-\epsilon$ \,\,\,& $0-\epsilon$\\
\end{tabular}\,\,
\end{center}
\vspace{4mm}

\noindent Between unconditional cooperators ($C$) and defectors ($D$), in agreement with the traditional formulation of a social dilemma \cite{santos_pnas06}, mutual cooperation yields the reward $R$, mutual defection leads to punishment $P$, and the mixed choice gives the cooperator the sucker's payoff $S$ and the defector the temptation $T$. Informed cooperators ($I_C$) and defectors ($I_D$), on the other hand, invest $\epsilon$ into knowing the strategy of their opponent and act accordingly. In particular, $I_C$ players refuse to be exploited by defectors, while $I_D$ players avoid being punished when encountering other defectors. We suppose that being informed conveys an advantage to $I_D$ players compared to $D$ players. Without increasing the number of parameters further, the simplest way to ensure this is that $D$s are still being punished if they meet with an $I_D$ player. For simplicity but without loss of generality, we may decrease the number of free parameters by introducing $T=1+r$, $R=1$, $P=-p$ and $S=-r$. Furthermore, the condition $0 < p < r$, namely that the punishment for mutual defection is less than the sucker's payoff and that the reward for mutual cooperation is smaller than the temptation to defect, yields $T>R>P>S$ and thus returns the most demanding and widely studied social dilemma --- the prisoner's dilemma game (cf. \cite{zimmermann_pre04, santos_prl05, gomez-gardenes_prl07, ohtsuki_prl07, fu_pre08b, poncela_epl09, fu_pre09, fu_jtb10, jiang_ll_pre10, antonioni_pone11, lee_s_prl11, tanimoto_pre12, hilbe_pnas13, santos_md_srep14}).

\begin{figure}
\begin{center}
\includegraphics[width = 4.0cm]{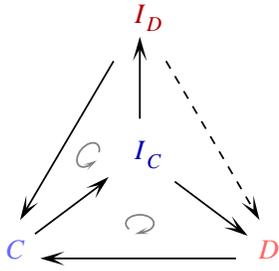}
\caption{\label{web}
The direction of invasion between the competing strategies (see arrows), where the two cyclically dominant triplets $I_C \to D \to C \to I_C$ and $I_C \to I_D \to C \to I_C$ form a defensive alliance against the external strategy (either $I_D$ or $D$).}
\end{center}
\end{figure}

With this definition of the payoffs, the directions of invasion between the competing strategies are as shown in Fig.~\ref{web}. An unconditional cooperator ($C$) will always beat an informed cooperator ($I_C$), simply because the latter has to bear the cost $\epsilon$. True to the formulation of a traditional social dilemma, an unconditional defector ($D$) will also beat $C$. This relation is reversed against $I_C$ players due to their ability to avoid exploitation, as long as $\epsilon < p$. The $I_C$ strategy is also superior to informed defectors ($I_D$) because although both have to bear $\epsilon$, $I_C$ players do benefit from mutual cooperation. Moreover, if we assume $\epsilon<p$ and thus also $\epsilon<r$ (because $r>p)$, $I_D$ beats both $D$ and $C$, respectively.

The relations between the strategies shown in Fig.~\ref{web} reveal the competition between two strategy triplets, namely $I_C \to D \to C \to I_C$ and $I_C \to I_D \to C \to I_C$. These two triplets form a so-called defensive alliance \cite{szabo_pre01b, szabo_jpa05, kim_bj_pre05} that protects them against the external fourth strategy that is not within the loop. For example, the $(I_C+I_D+C)$ triplet is able to prevent strategy $D$ from invading because both $I_C$ and $I_D$ beat $D$. Similarly, the $(I_C+D+C)$ prevents the invasion of $I_D$. Despite of these relatively simple and clear-cut relations, several open questions remain, such as which informed strategy is able to prevail and how high can the cost $\epsilon$ be, which defensive alliance is stable, or can the two triplets coexist?

Since pattern formation always plays a decisive role in structured populations \cite{szolnoki_jrsif14}, we proceed with an in-depth analysis of the proposed evolutionary game on a $L \times L$ square lattice with periodic boundary conditions. Unless stated differently, we use random initial conditions such that all four strategies are uniformly distributed across the lattice. We simulate the evolutionary process in accordance with the standard Monte Carlo simulation procedure comprising the following elementary steps. First, a randomly selected player $x$ acquires its payoff $\Pi_x$ by playing the game with all its four neighbors. Next, player $x$ randomly chooses one neighbor $y$, who then also acquires its payoff $\Pi_y$ in the same way as previously player $x$. Lastly, player $x$ adopts the strategy $s_y$ from player $y$ with a probability determined by the Fermi function $W(s_y \to s_x)={(1+\exp[(\Pi_x-\Pi_y)/K])}^{-1}$,
where $K=0.1$ quantifies the uncertainty related to the strategy adoption process \cite{szabo_pr07}. The selected value ensures that better-performing players are readily followed by their neighbors, although adopting the strategy of a player that performs worse is not impossible either. Note that the results remain qualitatively similar in a wide range within the $K<2$ interval. At very high $K$ values, however, the system arrives to the random imitation limit. Each full Monte Carlo step (MCS) gives a chance for every player to change its strategy once on average. During simulations, we have used lattices with up to $4000 \times 4000$ players to avoid undesired finite-size effects, and we have determined the stationary fractions of the strategies in the stationary state after sufficiently long relaxation times extending up to $4 \cdot 10^5$ MCS. In addition, we have repeated every run up to $100$ times with different initial conditions to further decrease statistical fluctuations.

\begin{figure}
\begin{center}
\includegraphics[width = 8.5cm]{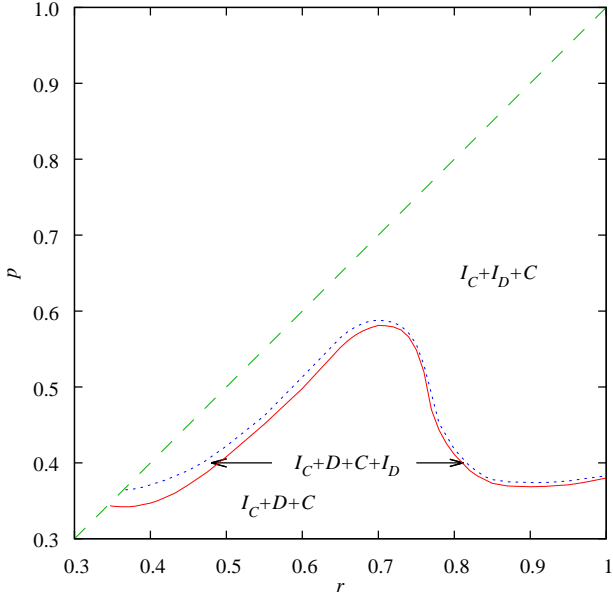}
\caption{\label{phase}Full $r-p$ phase diagram, as obtained for $\epsilon=0.3$. Solid red line denotes continuous phase transitions from the very narrow but stable $I_C+D+C+I_D$ phase to the stable $I_C+D+C$ phase, while the dotted blue line denotes reentrant continuous phase transitions to the stable $I_C+I_D+C$ phase. Dashed green line is the $p<r$ border.}
\end{center}
\end{figure}

\begin{figure}
\begin{center}
\includegraphics[width = 8.0cm]{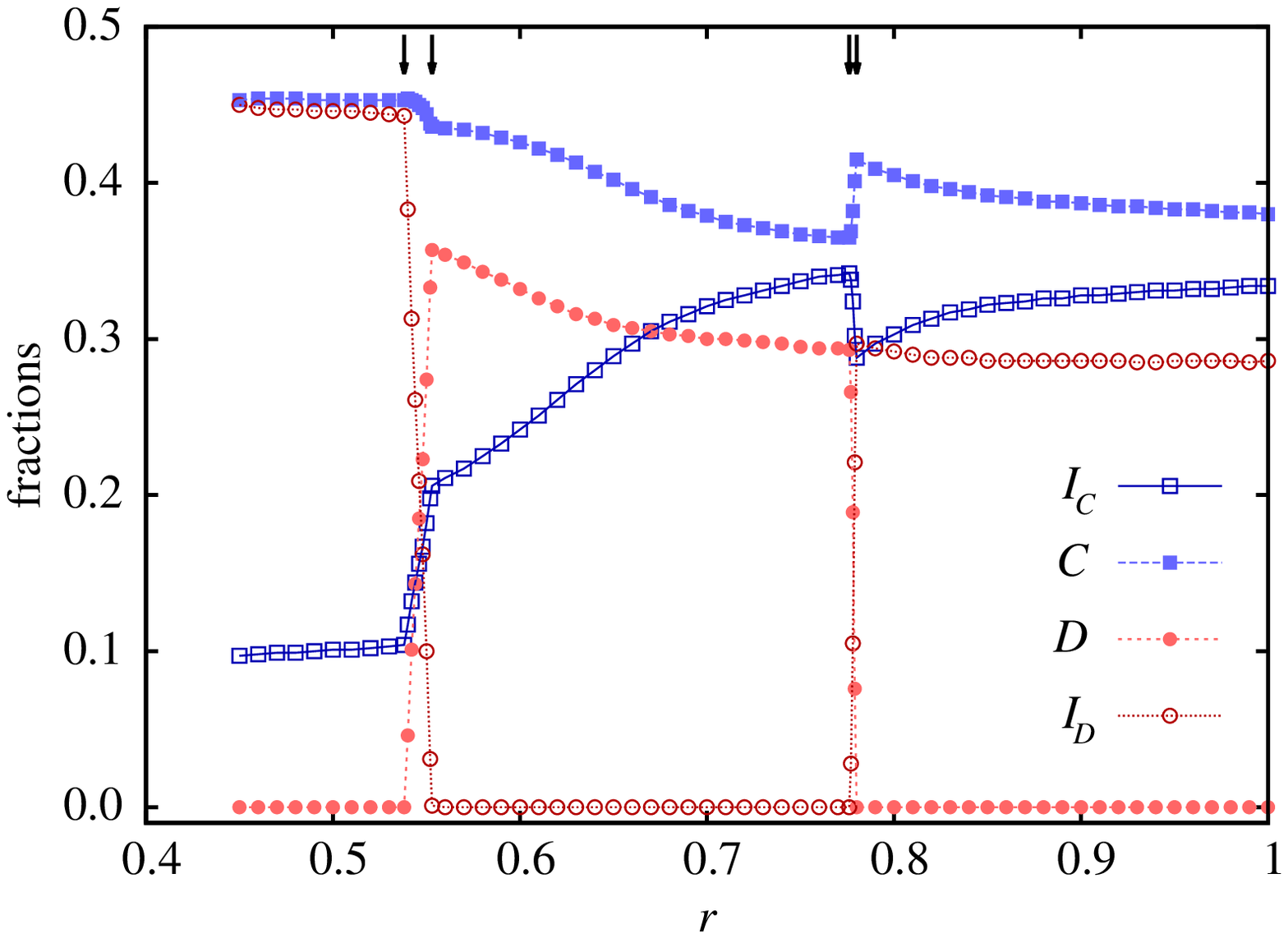}\\
\includegraphics[width = 8.0cm]{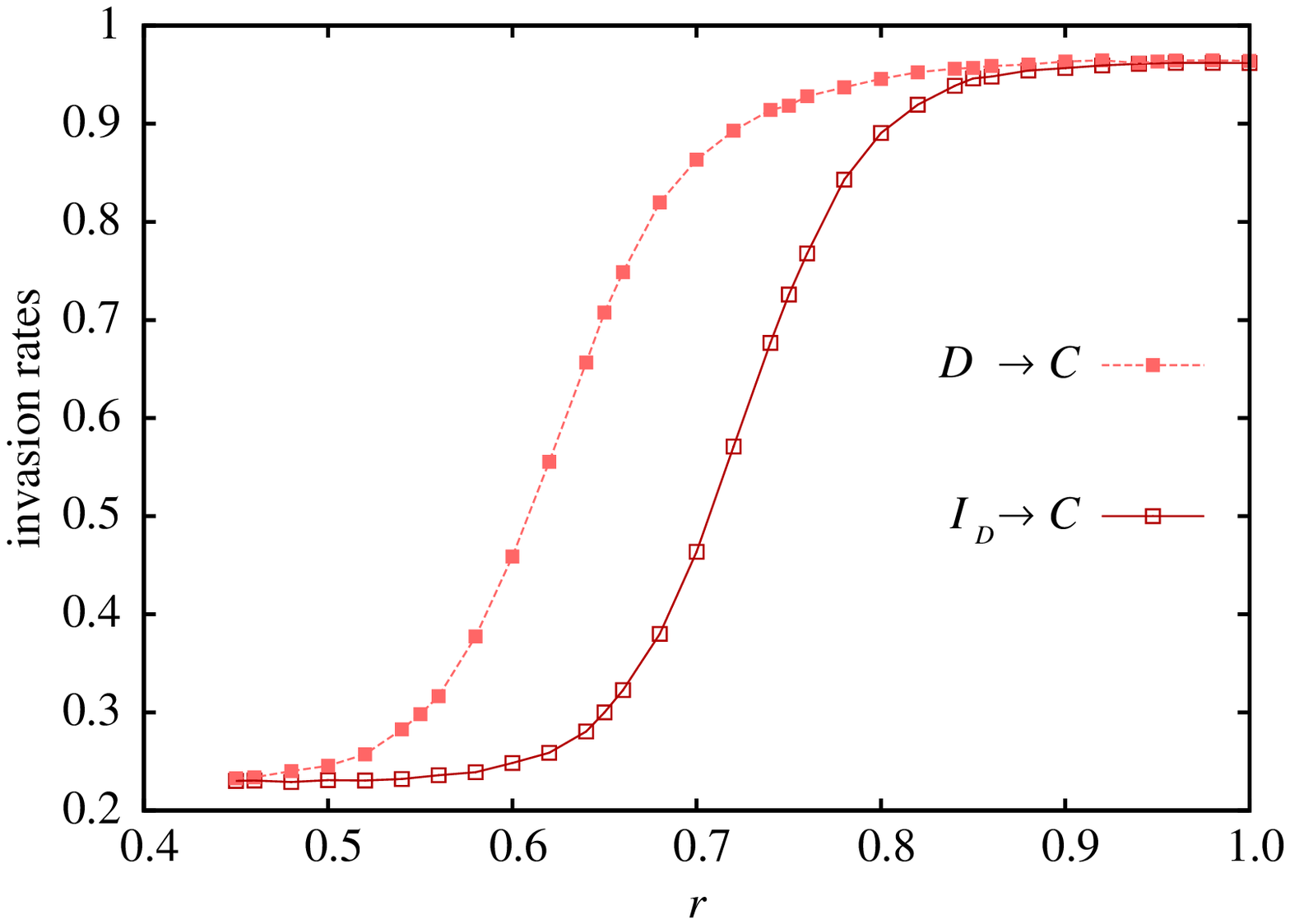}
\caption{\label{reentrant}Top panel shows a characteristic cross-section of the phase diagram in Fig.~\ref{phase}, as obtained for $p= 0.45$. Depicted are stationary fractions of the four competing strategies in dependence on $r$. After a succession of four continuous phase transitions, the system ends up in the same $I_C+I_D+C$ phase in which it started. The reentrant phase transition on both sides is preceded by a very narrow region (marked by arrows) where all four strategies coexist. Bottom panel shows the invasion rates between cooperator and defector strategies (see legend) in dependence on $r$, as obtained for $\epsilon=0.3$ and $p=0.45$. The hysteresis is crucial for the understanding of the reentrant phase transition to the $I_C+I_D+C$ phase (see main text for details).}
\end{center}
\end{figure}

In Fig.~\ref{phase}, we show the complete $r-p$ phase diagram of the system as obtained for a representative value of $\epsilon$. We would like to stress, however, that qualitatively similar behavior can be obtained for all relevant values of parameters $\epsilon$ or $K$. Several interesting solutions deserve to be highlighted. First, the $I_C$ strategy is always viable, even though in its absence defectors would easily eradicate unconditional cooperators. However, the later also become viable as second-order free-riders of $I_C$ (first-order free-riders are either $D$ or $I_D$ players). For low values of $p$ the $(I_C+D+C)$ triplet prevails and $I_D$ players die out, while for high values of $p$ the $(I_C+I_D+C)$ triplet wins on the expense of $D$. Interestingly, there is a very narrow dividing stripe where both triplets coexist, and where thus all four strategies survive. The dependence on $r$ is even more interesting. At an intermediate value of $p$, the system undergoes a succession of four consecutive phase transitions as $r$ increases, after which it ends in the same $I_C+I_D+C$ phase in which it started. This reentrant phase transition is illustrated in the top panel of Fig.~\ref{reentrant}, where both $I_C+I_D+C$ phases are separated by two very narrow $I_C+D+C+I_D$ stripes and a broader $I_C+D+C$ region in the middle. What is truly remarkable is that the competition between $D$ and $I_D$ (which exchange in the two triplets) is not directly affected by the value of $r$. What is more, four out of all the six relations in the food web are independent of $r$. While a reentrant phase transition has been reported before in the realm of predator-prey interactions due to rewiring \cite{han_pre09}, here it is due solely to strategic interactions.

An explanation for the unexpected behavior thus has to lie in the $D \to C$ and $I_D \to C$ invasion rates, which are directly affected by $r$. The bottom panel of Fig.~\ref{reentrant} reveals a monotonous dependence of the two invasion rates on $r$, whereby both curves start and terminate at an identical level. Thus, at both ends the two triplets have an equally fast internal dynamics, but since $I_D$ players beat $D$ players, the dominance belongs to the $(I_C+I_D+C)$ triplet. But for intermediate values of $r$, the $D \to C$ invasion is significantly more efficient than the $I_D \to C$ invasion, and the average rotation of strategies within the $(I_C+D+C)$ triplet is therefore much faster than in the $(I_C+I_D+C)$ triplet. Due to this difference in the internal dynamics, the $I_D$ players meet with their direct predator, the $I_C$ players (see the food web in Fig.~\ref{web}), more frequently. Because of this the faster rotating alliance is able to overcome the direct evolutionary disadvantage that $D$ players have against $I_D$ players, and in fact reverse the outcome of the competition between the two defensive alliances in favor of the $(I_C+D+C)$ triplet. Naturally, as soon as the two invasion rates meet at either the low or the high limit of $r$, the difference in the rotation speeds within the alliances disappears, and the original victor, the $(I_C+I_D+C)$ triplet, is restored.

\begin{figure*}[ht]
\begin{center}
\includegraphics[width = 4.5cm]{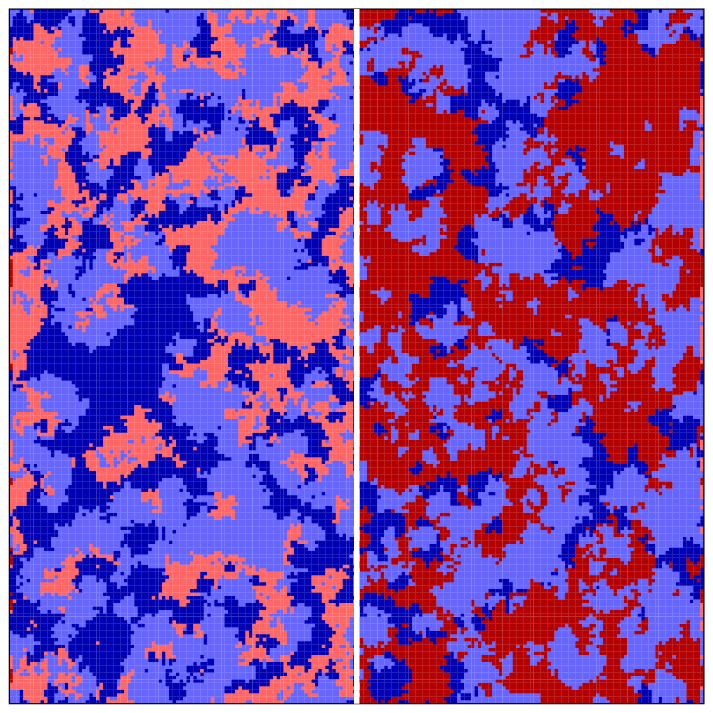}\includegraphics[width = 4.5cm]{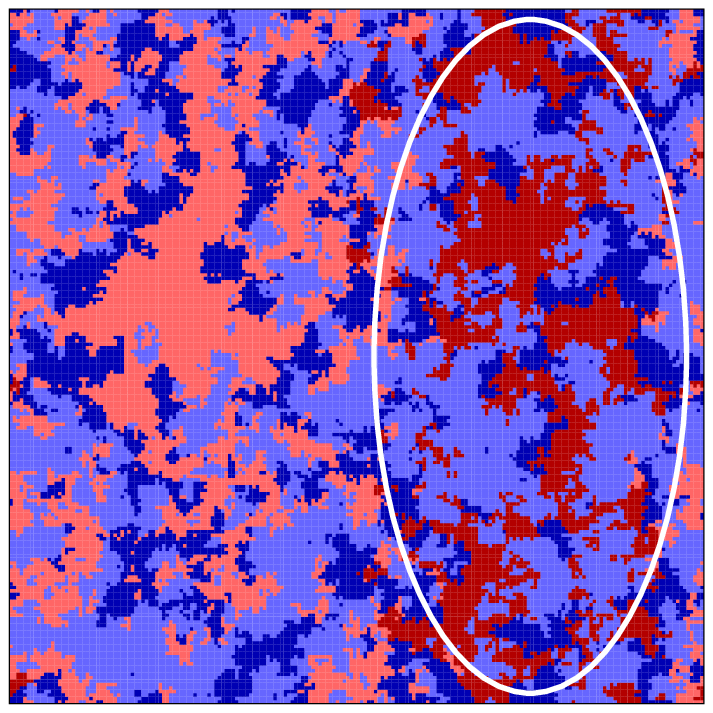}\includegraphics[width = 4.5cm]{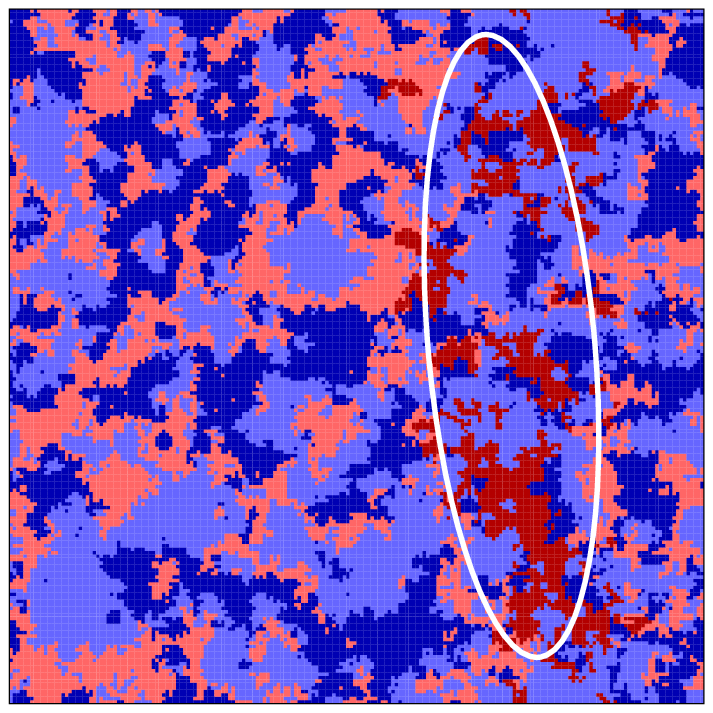}\includegraphics[width = 4.5cm]{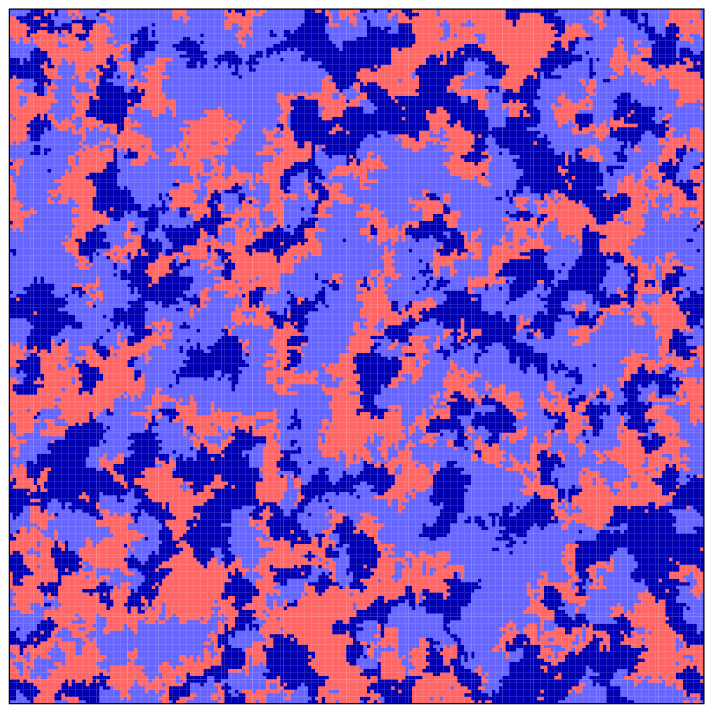}\\
\includegraphics[width = 4.5cm]{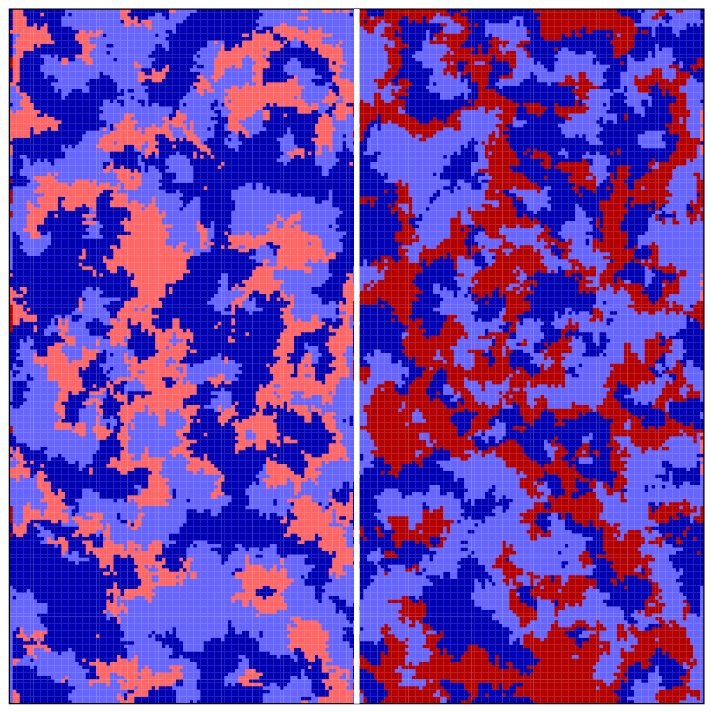}\includegraphics[width = 4.5cm]{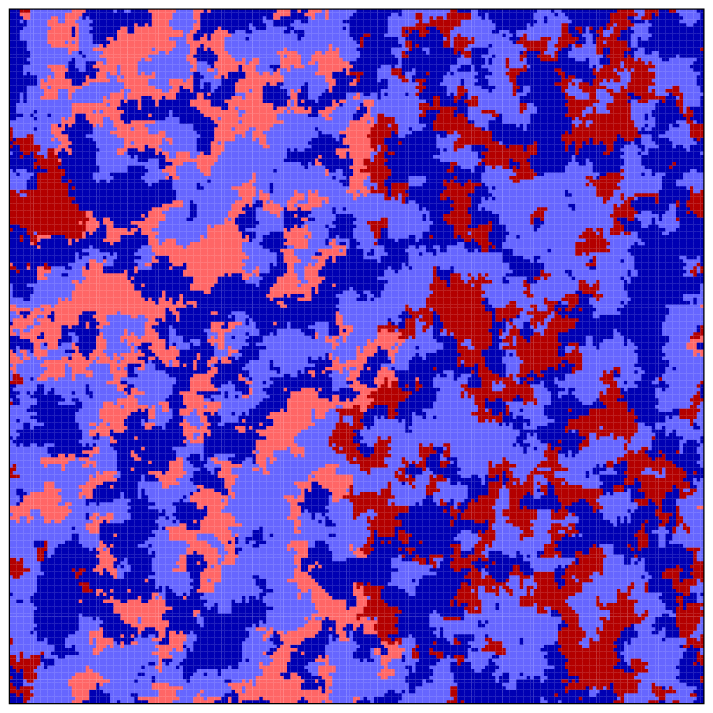}\includegraphics[width = 4.5cm]{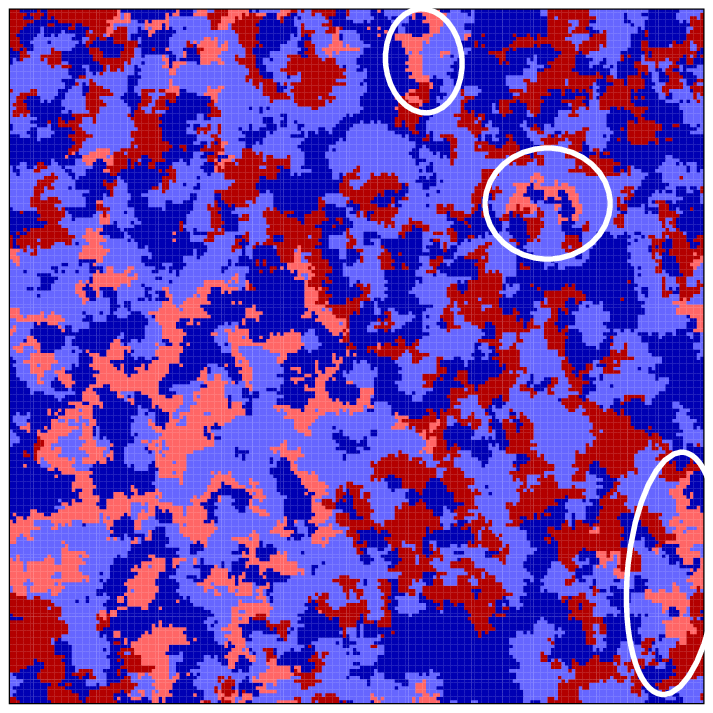}\includegraphics[width = 4.5cm]{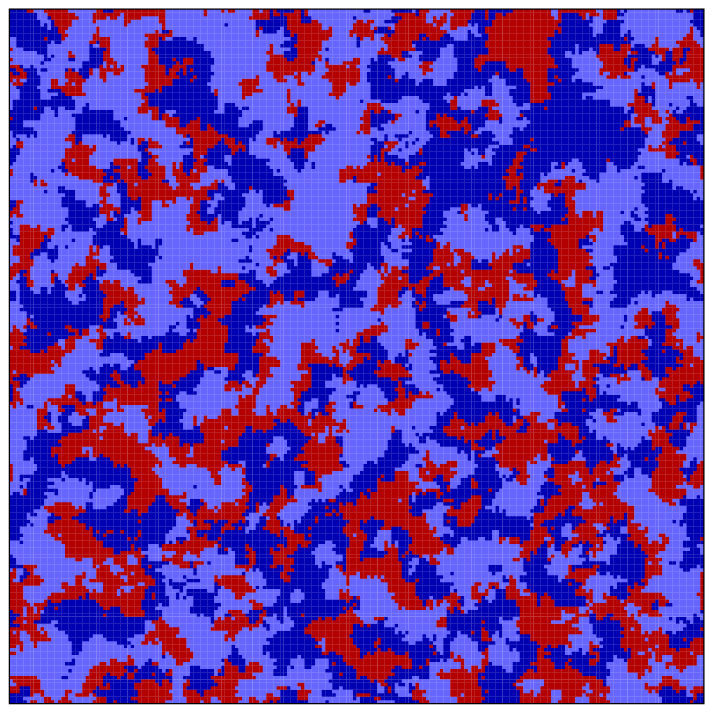}\\
\caption{\label{snaps}Consecutive snapshots of the square lattice, as obtained for $r=0.65$ (top row) and $r=0.95$ (bottom row) from a prepared initial state. During the relaxation period, each half of the lattice contained the strategies of one triplet only. After the characteristic patterns evolved (leftmost panels), the separating wall (vertical white line) was removed and the two defensive alliances could then start to compete with each other. In the top row, the $(I_C+D+C)$ triplet, which is depicted by dark blue, light blue, and light red colors (overall lighter), gradually compresses the $(I_C+I_D+C)$ triplet, which is depicted by dark blue, light blue, and dark red colors (overall darker). For convenience, the typical size of the shrinking alliance is encircled with a white ellipse. The snapshots were taken at $t=0$, $100$, $200$ and $500$ MCS from left to right. In the bottom row, the $(I_C+I_D+C)$ triplet gradually rises to dominance because $I_D$ players are superior to $D$ players. Accordingly, the domains of the $(I_C+D+C)$ triplet (encircled with white ellipses for clarity) disappear over time. Here the snapshots were taken at $t=0$, $100$, $1000$ and $4500$ MCS. Note that the evolutionary process in this case is almost ten times slower than in the top row. Other parameter values used in both cases are $L=200$, $p=0.4$ and $\epsilon=0.3$.}
\end{center}
\end{figure*}

As we have shown, when the $D \to C$ and $I_D \to C$ invasion rates are equal, the winner is decided by virtue of the direct dominance of $I_D$ players over $D$ players and thus the $(I_C+I_D+C)$ triplet wins. On the other hand, when the $D \to C$ invasion is significantly more efficient, then the faster average rotation of strategies within the $(I_C+D+C)$ triplet proves fatal for the $(I_C+I_D+C)$ triplet. These two different mechanisms that decide the competition between the two defensive alliances can also be appreciated through patterns that form on the square lattice, as illustrated in Fig.~\ref{snaps}. For clarity, we have split the square lattice in half and used prepared initial states to highlight the competition between the two triplets.

The top row of Fig.~\ref{snaps} depicts the evolution when the $(I_C+D+C)$ is the more dynamic triplet. It can be observed that the other triplet gradually decreases in size because its defensive capacity is compromised. To illustrate this monotonous decay, we have marked the typical size of the shrinking area by white ellipses during the intermediate stages of the evolutionary process. Although we only present a few snapshots, the difference in the rotation velocity within the two defensive alliances is still remarkably obvious. While the large majority of the typical pattern that characterizes the $(I_C+I_D+C)$ triplet (right side of the lattice) remains practically unchanged, the $(I_C+D+C)$ triplet (left side of the lattice) completes a whole strategy rotation during this time. Note that in the leftmost panel the majority of players within the $(I_C+D+C)$ alliance is in the cooperator state, in the second panel $D$ players abound, and then in the third panel the original fraction of $C+I_C$ players is again restored. Actually, such an ``oscillation'' is just a finite-size effect that is very useful for illustrative purposes, while in fact the fractions of the strategies are time-independent in the stationary state if a sufficiently large system size is used.

\begin{figure}
\begin{center}
\includegraphics[width = 8.0cm]{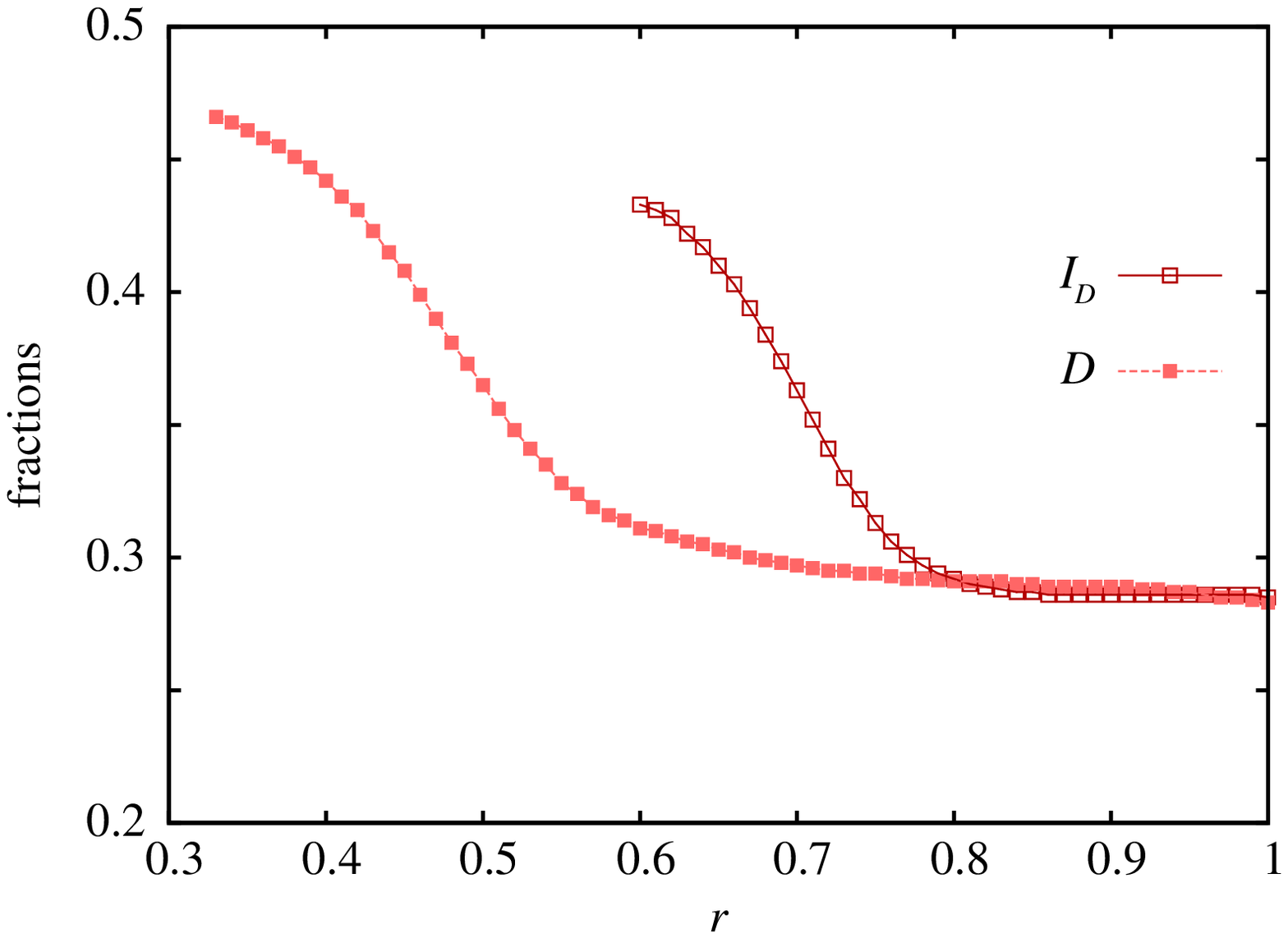}\\
\includegraphics[width = 8.0cm]{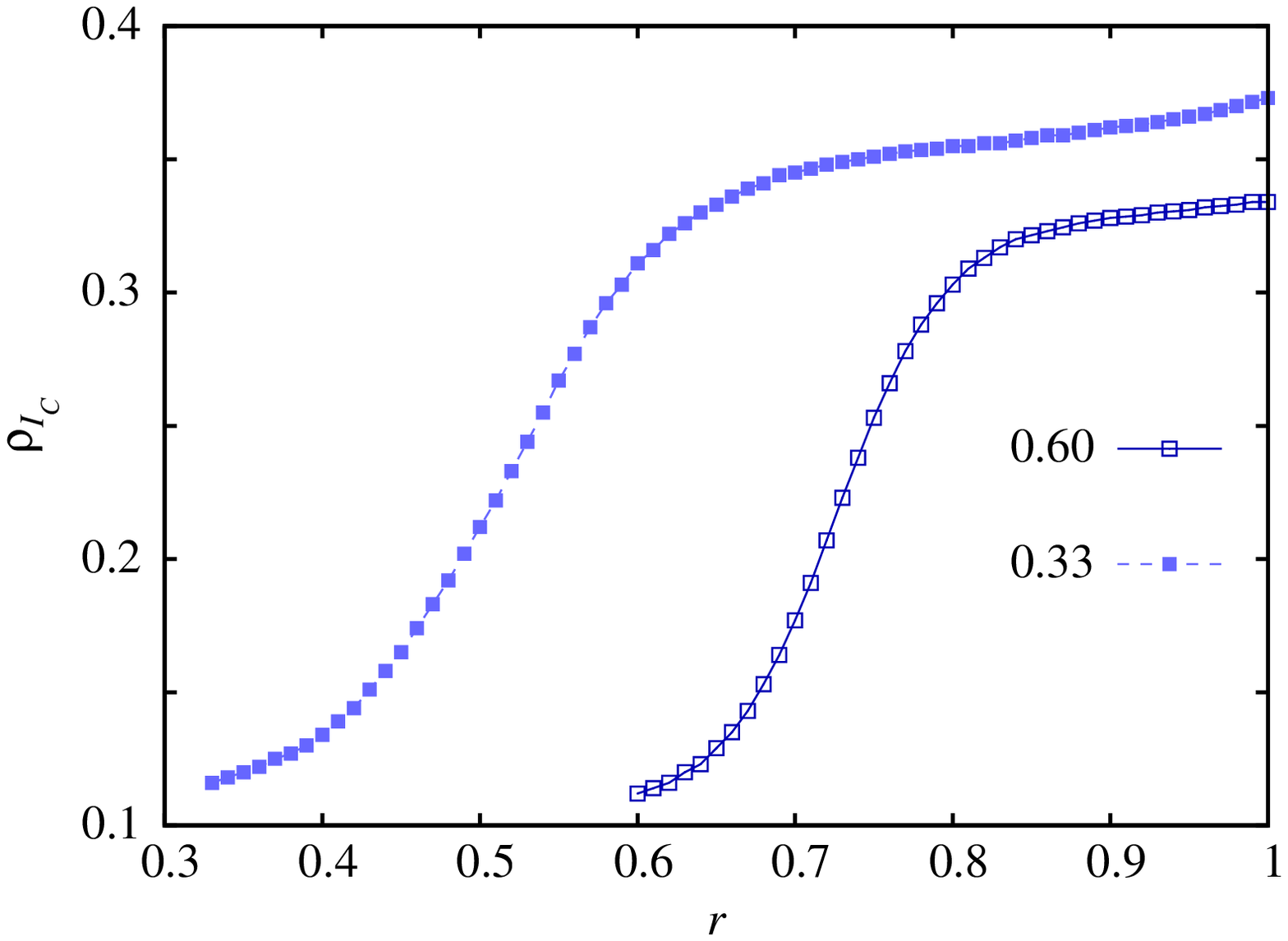}
\caption{\label{decay}Top panel shows the fraction of strategy $D$ in the $I_C+D+C$ phase at $p=0.33$ and the fraction of the $I_D$ strategy in the $I_C+I_D+C$ phase at $p=0.6$ in dependence on $r$, as obtained for $\epsilon=0.3$. The bottom panel shows the corresponding fraction of $I_C$ players ($\rho_{I_C}$) in the two triplets at the same two values of $p$ (see legend). Counterintuitively, due to cyclic dominance, as the temptation to defect increases, the fraction of defectors goes down while the fraction of informed cooperators goes up.}
\end{center}
\end{figure}

Conversely, the bottom row of Fig.~\ref{snaps} depicts the evolution when the strategies within the two triplets rotate equally fast. Accordingly, the pattern formation is significantly different from the one shown in the top row of Fig.~\ref{snaps}. Here the roles of strategies $I_D$ and $D$ in the two alliances are interchangeable --- both strategies beat strategy $C$ with practically the same efficiency. As a consequence, the patterns formed by the two triplets are very similar, and this not just approximately but quantitatively as well. If we compare the fractions of $D$ and $I_D$ players in the stable $I_C+D+C$ and $I_C+I_D+C$ phases around the right reentrant phase transition depicted in the top panel of Fig.~\ref{reentrant}, we can observe that they are practically equivalent (just switch values). Indeed, the superiority of the $I_D$ strategy manifests only when it is faced with strategy $D$. Because of this ``individual'' competition, the evolutionary process is much slower than in the top row (note that it takes ten times longer to reach the single-triple stationary state). Another more subtle consequence of identically fast average rotations of strategies within the two competing triplets is that there is a good chance for a lucky $D$ player to emerge and spread efficiently on the expense of $C$ players. In doing so, regions of the square lattice that already belonged to the $(I_C+I_D+C)$ triplet may be temporarily lost to the $(I_C+D+C)$ triplet. We have marked such unexpected patches of the $(I_C+D+C)$ alliance by white ellipses in the third panel of the bottom row of Fig.~\ref{snaps}.

An unexpected but important consequence of cyclic dominance in the studied system is also the peculiar dependence of the fractions of individual strategies within the two defensive alliances on $r$, which we illustrate in Fig.~\ref{decay}. Based on the payoff matrix, we would expect that increasing values of $r$ will unambiguously support both defector strategies and at the same time weaken unconditional cooperators, as is the case in the traditional two-strategy prisoner's dilemma game. But as results presented in Fig.~\ref{decay} show, exactly the opposite happens regardless of which defensive alliance is dominant. This counterintuitive behavior is a beautiful consequence of cyclic dominance, where the direct support for a particular strategy will actually promote its ``superior'' in the food web. Such evolutionary outcomes are common in rock-paper-scissors-like systems, where feeding a prey often promotes the predator \cite{frean_prsb01, dobrinevski_pre14, ni_x_pre10, wang_wx_pre10b, rulands_pre13, vukov_pre13, laird_jtb15, groselj_pre15}. In the studied social dilemma with informed strategies this is exactly what happens, since increasing the value of $r$ will directly support $D$ and $I_D$ strategies, which however are both prey to the $I_C$ strategy (see the food web in Fig.~\ref{web}). As demonstrated in the bottom panel of Fig.~\ref{decay}, indeed the fraction of $I_C$ goes up in both phases as $r$ increases. An additional consequence of increasing the value of $r$ is that it lowers the fitness of $C$ players. Since the latter act as second-order free-riders of $I_C$ players, and are at the same time prey to both $D$ and $I_D$ strategies, this also promotes the viability of informed cooperators.

Before summarizing our findings, we briefly discuss the role of the information cost $\epsilon$. Throughout this work, we have considered $\epsilon=0.3$, as it yields the most representative system dynamics. By increasing the value of $\epsilon$, the informed strategies become weaker. While the $I_D \to D$ relation still holds, the intensity of this invasion decreases. Consequently, the winning mechanism of the $(I_C+I_D+C)$ triplet is negatively affected, and the region where the faster rotating $(I_C+D+C)$ triplet is able to compensate the $I_D \to D$ relation expands. For $\epsilon>0.64$ the $I_C+I_D+C$ alliance disappears completely. On the other hand, lower $\epsilon$ values intensify the $I_D \to D$ invasion, and accordingly, the region where the $I_C+I_D+C$ triplet is dominant expands on the expense of the $I_C+D+C$ triplet. Nevertheless, the faster rotating $I_C+D+C$ alliance can defend its stability throughout the small $\epsilon$ region.

In summary, we have proposed and studied an evolutionary social dilemma game with informed strategies. We have identified elementary relations between the four competing strategies in the governing food web, which revealed the existence of two three-strategy defensive alliances. By means of systematic Monte Carlo simulations, we have revealed key determinants of stability of the two competing triplets, and we have observed fascinating solutions that are driven by pattern formation and cyclic dominance. We have shown that a direct evolutionary advantage of a strategy within a defensive alliance can be compensated by the other alliance through a faster internal rotation of its strategies. Thus, even though in the food web the informed defectors are superior to unconditional defectors, the alliance whose defense relies on the weaker strategy can still prevail. In particular, we have shown that the competition between a direct food-web-based evolutionary advantage and an evolutionary advantage that is rooted in the spatiotemporal dynamics of a defensive alliance gives rise to a reentrant phase transition. In-between the two stable three-strategy phases that form the reentrant pair, we have also identified a very narrow region of coexistence of both defensive alliances, which emerges as a consequence of a delicate equilibrium of the two competing mechanisms.

When approaching more realistic conditions to study the evolution of cooperation \cite{capraro_pone14, hauser_n14}, the emergence of cyclic dominance becomes more and more likely. Our research highlights that the final outcome in such cases depends sensitively not just on the individual relations between the competitors as determined by payoff elements, but in an equally strong manner also on the dynamical properties of alliances. This in turn makes interventions aimed as steering such complex systems risky and to a large degree unpredictable.

\begin{acknowledgments}
This research was supported by the Hungarian National Research Fund (Grant K-101490), the Slovenian Research Agency (Grant P5-0027), and the Deanship of Scientific Research, King Abdulaziz University (Grant 76-130-35-HiCi).
\end{acknowledgments}


\begin{thebibliography}{10}
\expandafter\ifx\csname url\endcsname\relax\def\url#1{\texttt{#1}}\fi

\bibitem{szabo_pr07}
\Name{Szab{\'o} G. \and F{\'a}th G.} \REVIEW{Phys. Rep.}{446}{2007}{97}.

\bibitem{santos_jtb12}
\Name{Santos F.~C., Pinheiro F., Lenaerts T. \and Pacheco J.~M.} \REVIEW{J.
  Theor. Biol.}{299}{2012}{88}.

\bibitem{nowak_jtb12}
\Name{Nowak M.~A.} \REVIEW{J. Theor. Biol.}{299}{2012}{1}.

\bibitem{perc_bs10}
\Name{Perc M. \and Szolnoki A.} \REVIEW{BioSystems}{99}{2010}{109}.

\bibitem{rand_tcs13}
\Name{Rand D.~A. \and Nowak M.~A.} \REVIEW{Trends in Cognitive Sciences}{17}{2013}{413}.

\bibitem{pacheco_plrev14}
\Name{Pacheco J.~M., Vasconcelos V.~V. \and Santos F.~C.} \REVIEW{Physics of
  Life Reviews}{11}{2014}{573}.

\bibitem{mestertong_01}
\Name{Mesterton-Gibbons M.} \Book{An Introduction to Game-Theoretic Modelling,
  2nd Edition} (American Mathematical Society, Providence, RI) 2001.

\bibitem{nowak_06}
\Name{Nowak M.~A.} \Book{Evolutionary Dynamics} (Harvard University Press,
  Cambridge, MA) 2006.

\bibitem{sigmund_10}
\Name{Sigmund K.} \Book{The Calculus of Selfishness} (Princeton University
  Press, Princeton, NJ) 2010.

\bibitem{lugo_srep15}
\Name{Lugo H. \and San~Miguel M.} \REVIEW{Sci. Rep.}{5}{2015}{7776}.

\bibitem{wu_zx_pre14}
\Name{Wu Z.-X. \and Yang H.-X.} \REVIEW{Phys. Rev. E}{89}{2014}{012109}.

\bibitem{nowak_s06}
\Name{Nowak M.~A.} \REVIEW{Science}{314}{2006}{1560}.

\bibitem{neumann_44}
\Name{von Neumann J. \and Morgenstern O.} \Book{Theory of Games and Economic
  Behaviour} (Princeton University Press, Princeton, NJ) 1944.

\bibitem{macy_pnas02}
\Name{Macy M.~W. \and Flache A.} \REVIEW{Proc. Natl. Acad. Sci. USA}{99}{2002}{7229}

\bibitem{rustagi_s10}
\Name{Rustagi D., Engel S. \and Kosfeld M.} \REVIEW{Science}{330}{2010}{961}

\bibitem{szolnoki_pre12}
\Name{Szolnoki A. \and Perc M.} \REVIEW{Phys. Rev. E}{85}{2012}{026104}

\bibitem{antonioni_pone14}
\Name{Antonioni A., Cacault M.~P., Lalive R. \and Tomassini M.} \REVIEW{PLoS
  ONE}{9}{2014}{e110788}.

\bibitem{santos_pnas06}
\Name{Santos F.~C., Pacheco J.~M. \and Lenaerts T.} \REVIEW{Proc. Natl. Acad.
  Sci. USA}{103}{2006}{3490}.

\bibitem{zimmermann_pre04}
\Name{Zimmermann M.~G., Egu{\'{\i}}luz V.~M. \and San~Miguel M.} \REVIEW{Phys.
  Rev. E}{69}{2004}{065102(R)}.

\bibitem{santos_prl05}
\Name{Santos F.~C. \and Pacheco J.~M.} \REVIEW{Phys. Rev. Lett.}{95}{2005}{098104}.

\bibitem{gomez-gardenes_prl07}
\Name{G{\'o}mez-Garde{\~n}es J., Campillo M., Flor{\'{\i}}a L.~M. \and Moreno
  Y.} \REVIEW{Phys. Rev. Lett.}{98}{2007}{108103}.

\bibitem{ohtsuki_prl07}
\Name{Ohtsuki H., Nowak M.~A. \and Pacheco J.~M.} \REVIEW{Phys. Rev. Lett.}{98}{2007}{108106}.

\bibitem{fu_pre08b}
\Name{Fu F., Hauert C., Nowak M.~A. \and Wang L.} \REVIEW{Phys. Rev. E}{78}{2008}{026117}.

\bibitem{poncela_epl09}
\Name{Poncela J., G{\'o}mez-Garde{\~n}es J., Flor{\' \i}a L.~M., Moreno Y. \and
  S{\'a}nchez A.} \REVIEW{EPL}{88}{2009}{38003}.

\bibitem{fu_pre09}
\Name{Fu F., Wu T. \and Wang L.} \REVIEW{Phys. Rev. E}{79}{2009}{036101}.

\bibitem{fu_jtb10}
\Name{Fu F., Nowak M.~A. \and Hauert C.} \REVIEW{J. Theor. Biol.}{266}{2010}{358}.

\bibitem{jiang_ll_pre10}
\Name{Jiang L.-L., Wang W.-X., Lai Y.-C. \and Wang B.-H.} \REVIEW{Phys. Rev. E}{81}{2010}{036108}.

\bibitem{antonioni_pone11}
\Name{Antonioni A. \and Tomassini M.} \REVIEW{PLoS ONE}{6}{2011}{e25555}.

\bibitem{lee_s_prl11}
\Name{Lee S., Holme P. \and Wu Z.-X.} \REVIEW{Phys. Rev. Lett.}{106}{2011}{028702}.

\bibitem{tanimoto_pre12}
\Name{Tanimoto J., Brede M. \and Yamauchi A.} \REVIEW{Phys. Rev. E}{85}{2012}{032101}.

\bibitem{hilbe_pnas13}
\Name{Hilbe C., Nowak M. \and Sigmund K.} \REVIEW{Proc. Natl. Acad. Sci. USA}{110}{2013}{6913}.

\bibitem{santos_md_srep14}
\Name{Santos M., Dorogovtsev S.~N. \and Mendes J. F.~F.} \REVIEW{Sci. Rep.}{4}{2014}{4436}.

\bibitem{szabo_pre01b}
\Name{Szab{\'o} G. \and Cz{\'a}r{\'a}n T.} \REVIEW{Phys. Rev. E}{64}{2001}{042902}.

\bibitem{szabo_jpa05}
\Name{Szab{\'o} G.} \REVIEW{J. Phys. A: Math. Gen.}{38}{2005}{6689}.

\bibitem{kim_bj_pre05}
\Name{Kim B.~J., Liu J., Um J. \and Lee S.-I.} \REVIEW{Phys. Rev. E}{72}{2005}{041906}.

\bibitem{szolnoki_jrsif14}
\Name{Szolnoki A., Mobilia M., Jiang L.-L., Szczesny B., Rucklidge A.~M. \and
  Perc M.} \REVIEW{J. R. Soc. Interface}{11}{2014}{20140735}.

\bibitem{han_pre09}
\Name{Han S.-G., Park S.-C. \and Kim B.-J.} \REVIEW{Phys. Rev. E}{79}{2009}{066114}.

\bibitem{frean_prsb01}
\Name{Frean M. \and Abraham E.~D.} \REVIEW{Proc. R. Soc. Lond. B}{268}{2001}{1323}.

\bibitem{dobrinevski_pre14}
\Name{Dobrinevski A., Alava M., Reichenbach T. \and Frey E.} \REVIEW{Phys. Rev.
  E}{89}{2014}{012721}.

\bibitem{ni_x_pre10}
\Name{Ni X., Wang W.-X., Lai Y.-C. \and Grebogi C.} \REVIEW{Phys. Rev. E}{82}{2010}{066211}.

\bibitem{wang_wx_pre10b}
\Name{Wang W.-X., Lai Y.-C. \and Grebogi C.} \REVIEW{Phys. Rev. E}{81}{2010}{046113}.

\bibitem{rulands_pre13}
\Name{Rulands S., Zielinski A. \and Frey E.} \REVIEW{Phys. Rev. E}{87}{2013}{052710}.

\bibitem{vukov_pre13}
\Name{Vukov J., Szolnoki A. \and Szab{\'o} G.} \REVIEW{Phys. Rev. E}{88}{2013}{022123}.

\bibitem{laird_jtb15}
\Name{Laird R.~A. \and Schamp B.~S.} \REVIEW{J. Theor. Biol.}{365}{2015}{149}.

\bibitem{groselj_pre15}
\Name{Gro{\v s}elj D., Jenko F. \and Frey E.} \REVIEW{Phys. Rev. E}{91}{2015}{033009}.

\bibitem{capraro_pone14}
\Name{Capraro V., Smyth C., Mylona K. \and Niblo G.~A.} \REVIEW{PLoS ONE}{9}{2014}{e102881}.

\bibitem{hauser_n14}
\Name{Hauser O.~P., Rand D.~G., Peysakhovich A. \and Nowak M.~A.}
  \REVIEW{Nature}{511}{2014}{220}.

\end{thebibliography}
\end{document}